\documentclass[aps,twocolumn,showpacs,english]{revtex4}

\usepackage[T1]{fontenc}
\usepackage[latin9]{inputenc}
\usepackage{textcomp}
\usepackage{amsmath}
\usepackage{graphicx}
\usepackage{amssymb}
\usepackage{babel}
\begin{document}

\title{Dark exciton optical spectroscopy \\ of a semiconducting quantum dot
 embedded in a nanowire}

\author{G. Sallen$^{1}$, A. Tribu$^{2}$, T. Aichele$^{1}$, R. Andr\'{e}$^{1}$, L. Besombes$^{1}$, C. Bougerol$^{1}$, S. Tatarenko$^{1}$, K. Kheng$^{2}$, and J. Ph.~Poizat$^{1}$}

\affiliation{CEA-CNRS-UJF group 'Nanophysique et Semiconducteurs',\\
$^1$ Institut N\'{e}el, CNRS - Universit\'{e} Joseph Fourier, 38042 Grenoble, France, \\
$^2$ CEA/INAC/SP2M, 38054 Grenoble, France}

\begin{abstract}
 Photoluminescence
of a single CdSe quantum dot embedded in a ZnSe nanowire has been investigated. It has been found that the dark exciton has a strong influence on the optical properties.
The most visible influence is the
 strongly reduced excitonic emission compared to the biexcitonic one.
Temperature dependent lifetime measurements have allowed us to measure a large splitting of $\Delta E = 6 $ meV between the dark and the bright exciton as well as the spin flip rates between these two states.

\end{abstract}

\pacs{78.67.Lt, 78.55.Et}

\maketitle

Specific growth techniques developped about a decade ago have led to semiconducting nanowires (NW) that have attracted a great deal of interest since then. Their potential applications includes nanoelectronics \cite{Duan01,Lu,The03}, optoelectronics (light emitting diodes  \cite{Konenkamp,Kim}, nanolasers \cite{Duan03}),  thermoelectrical energy conversion  \cite{Hochbaum}, and biological or chemical sensors \cite{Cui}.

By changing the material composition during the growth it is possible to
change the chemical composition \cite{Gud02,Bjork}  along the longitudinal or radial directions. This enables the fabrication of well controlled  1D nanoscale  heterostructures \cite{Bjork}. For example, as shown in this work, it is possible to insert a slice of a low band gap semiconductor within a high band gap NW and realize a light emitting quantum dot (QD) \cite{borgstrom,Tribu}. The absence of a wetting layer offers a better confinement compared to self-assembled QDs. This has allowed our group to produce  single photon at high temperature ($220$ K) \cite{Tribu}.
Furthermore,  NW based heterostructures, being much less limited by lattice mismatches, greatly widen the possible materials combinations and
enable well controlled stacking of several QDs in a single NW,   offering interesting possibilities for quantum information processing \cite{Simon}.

In a QD, lowest energy excitons are the combination of an electron (spin $\pm 1/2$) and a heavy hole (spin $\pm 3/2$). This results in two different energy levels of spin $\pm 1$ and spin $\pm 2$. The spin $\pm 1$ states are optically connected to the zero spin empty dot state and called the bright exciton. The low energy spin $\pm 2$ states are called the dark exciton because they are not optically
 active. Indeed a  photon is a spin 1 particle that can not carry away 2 quanta of angular momentum.

In a previous work we have
 performed a thorough  spectroscopic analysis   of a single CdSe QD embedded in a ZnSe NW by using photon correlation spectroscopy \cite{correlation}. We have  identified unambiguously the exciton, biexciton, and charged exciton lines using cross-correlations and obtained information on the charging dynamics of this QD. In order to fit the various correlation functions we had to include a dark exciton in the model. The most apparent manifestation of the dark exciton is the large predominance of the biexciton line with respect to the exciton line above saturation.

In this paper we have performed temperature dependent lifetime measurements on this CdSe/ZnSe NW. By fitting these data with a model involving an acoustic phonon bath \cite{Labeau} we are able to extract the value of the dark and bright exciton energy splitting $\Delta E$ and the spin flip rates between these two states.

ZnSe NWs are grown by Molecular Beam Epitaxy (MBE) in the Vapour-Liquid-Solid (VLS) growth
mode catalysed with gold particles on a Si substrate.
In order to fabricate QDs, a small region of CdSe is inserted in  the ZnSe NW.
This is done by interrupting the ZnSe growth,
changing to CdSe for a short time and growing  ZnSe again \cite{Tribu}.
From the CdSe growth time, the height of
the CdSe slice is estimated to be between 1.5 and 4 nm .
The diameter (around $10$ nm) is of
the order of the bulk exciton Bohr diameter for CdSe ($2a_B=11$ nm). This means that the carriers in the CdSe QD are
in the strong confinement regime. Details on the growth of the ZnSe NWs can be found in  \cite{Aichele}.
For the study of single
NWs, the sample is sonicated in methanol, so that NWs broke off the substrate into
the solution.  Droplets of this
solution are then deposited on a Si substrate, and a
 low density of individual NWs is obtained after evaporation.

The experimental set-up is a standard  microphotoluminescence set-up. The
samples are mounted on a XYZ piezo motor system in a He flow cryostat allowing experiments at a temperature of
4 K.  Time resolved measurements are performed by illuminating the sample at a wavelength of $\lambda=440$ nm
with a frequency doubled Ti:sapphire laser operating at $\lambda=880$ nm with pulse duration of 1 ps and a repetition rate of 80 MHz.
Continuous wave excitation is provided by a $405$ nm continuous-wave (CW) diode laser.
The ZnSe band gap ($2.7$ eV) corresponds to a wavelength of $\lambda=460$ nm so that both excitation lasers create free carriers in the barrier.

The excitation light is illuminating the sample
via a microscope objective of numerical aperture $NA=0.65$ located in the cryostat.
The NW emission is
 collected by the same objective  and
 sent  to a monochromator
(1200 grooves/mm grating,  50 cm focal length).
The monochromator has a switchable mirror inside
that can direct the luminescence either onto a charge coupled device (CCD) camera for the
measurement of the microphotoluminescence ($\mu$PL) spectra or through the exit slit towards
a low jitter (40 ps) Avalanche Photodiode (APD).
The APD sends electrical pulses into
a time-correlated single photon module that record the arrival time of the photon with respect to the the laser pulse. The overall
temporal resolution of our set-up is essentially limited by the
jitter of the APD and the dispersion of the monochromator
grating. It was measured by sending the
 1 ps  laser pulses in the monochromator and a full width at half maximum
 of 70 ps was obtained.

\begin{figure}
  \resizebox{0.45\textwidth}{!}{\includegraphics{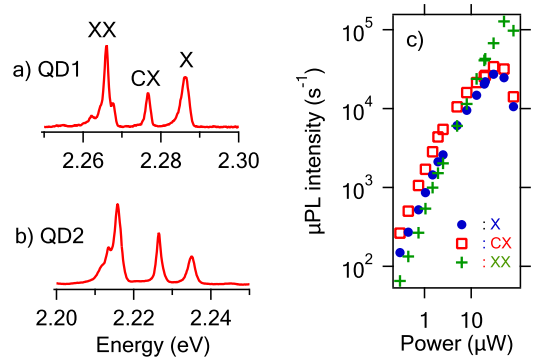}}
 \caption{Above saturation microphotoluminescence spectra of two different QDs. a) QD1,
 b) QD2 (excitation power P =60 $\mu$W).
 c) QD2 line intensities as a function of excitation power
}
 \label{fig:spectraD}
\end{figure}

Typical spectra are shown in fig. \ref{fig:spectraD}.
 A comparison with relative energy positions of known emission lines in spectra of self-assembled CdSe/ZnSe QDs \cite{Turck,Patton} suggests that these lines
 correspond to the exciton (X), the biexciton (XX) and the charged exciton (CX).  The X-CX and X-XX energy splitting are found around 10 meV (20 meV) as compared to 15-22 meV (19-26 meV) for self-assembled QDs. The mean excitonic energy is also similar ($2.25 \pm 0.08$ eV) as compared to $2.45 \pm 0.2$ eV  for self-assembled CdSe/ZnSe QDs \cite{data}.
Unambiguous proof for the assignment of these lines has been given using photon correlation spectroscopy \cite{correlation}.

\begin{figure}
 \resizebox{0.2\textwidth}{!}{\includegraphics{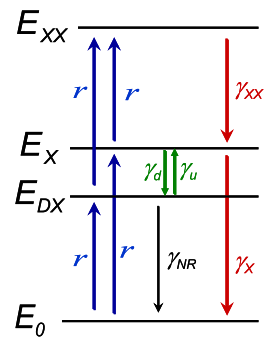}}
 \caption{Level scheme including the empty dot ($E_0$), the dark exciton ($E_{DX}$), the bright exciton ($E_{X}$) and the biexciton ($E_{XX}$).
 The various rates between the different level are indicated. The fixed values used in the model are $\gamma_X=1.4$ ns$^{-1}$, $\gamma_{XX}=2.5$ ns$^{-1}$, $\gamma_{CX}=1.7$ ns$^{-1}$. The parameter $r$ is the pumping rate.
  The other parameters are temperature dependent and are given in the text and in fig. \ref{fig:gammaSP}.}
 \label{fig:level_schemeD}
\end{figure}

The ratio between charged  and  neutral QD luminescence is varying from dot to dot. We have also observed that increasing the temperature tends to neutralize the QD.
The large linewidths have been attributed to spectral diffusion \cite{these}.
The most conspicuous feature is  that  the $\mu$PL intensity  of the XX line at saturation is always a lot larger than that of the X line as it is shown in the power dependence of the different lines (Fig. \ref{fig:spectraD} c)).
This effect is the signature of a strong storage effect on the dark exciton (DX) state. The DX state reduces the luminescence of the X line owing to the leakage from the bright to the dark exciton  but the DX state remains an efficient intermediate state for populating the XX state \cite{Reischle}.
The QD photoluminescence properties are well described by a set of rate equations including the bright and dark exciton,  and the biexciton as represented in fig. \ref{fig:level_schemeD}.

 The population transfer between the bright and the dark exciton states is  governed by the two temperature dependent  rates
 $\gamma_d$ and $\gamma_u$ (see fig. \ref{fig:level_schemeD}). We assume that these transitions are assisted by acoustic phonons whose energy matches the X-DX energy splitting $\Delta E$.
 At a temperature T the number $N$ of acoustic phonons per quantum state of energy $\Delta E$  is given by  the Bose-Einstein statistics and reads
 \begin{equation}
 N=\frac{1}{\exp(\Delta E/k_BT)-1}.
 \label{eq:BES}
 \end{equation}
 The downward transition rate $\gamma_d$ from the X to the DX state corresponds to the spontaneous and stimulated emission of a phonon, whereas the upward rate $\gamma_u$ corresponds to the absorption of a phonon \cite{Labeau}.
They are given by
 \begin{eqnarray}
 \gamma_d &=& (N+1)\gamma_0  , \nonumber \\
 \gamma_u &=& N \gamma_0 ,
 \label{eq:gamma}
 \end{eqnarray}
 where $\gamma_0$ is the zero temperature downward rate ($N=0$).

\begin{figure}
  \resizebox{0.5\textwidth}{!}{\includegraphics{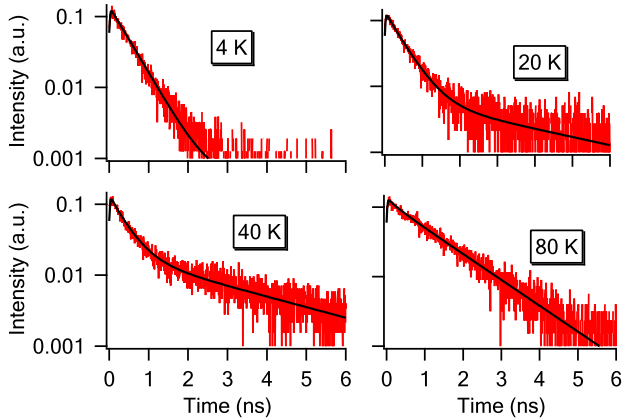}}
 \caption{ Decay of the X line emission of QD1 for different temperatures. The pumping power is well below saturation ($r\ll \gamma_X$) so that the XX state is almost not populated.}
 \label{fig:lifetimes}
\end{figure}

 We have performed time resolved photoluminescence of the X state of QD1 at different temperatures. The results are  presented in fig. \ref{fig:lifetimes}.
 The decay time of the X level depends not only on the radiative decay rate $\gamma_X$ but also on the temperature dependent $\gamma_d$ and $\gamma_u$ rates between the bright and dark excitons.
 At low temperature ($T=4 K$), the luminescence exhibits a fast monoexponential decay with a time scale of the order of $1/(\gamma_X + \gamma_d)$ corresponding  to the radiative decay and the leakage towards the DX state. For intermediate temperatures ($T=20$ K and $T=40$ K), the fast decay is still present and there is the apparition of a slow time scale corresponding to
 the thermally activated reloading of the X state from the long lived DX state. For higher temperature ($T=80$ K), the reloading from the dark to the bright exciton becomes even more efficient, and the decay appears as monoexponential with a time scale intermediate between the previous slow and  fast time scales.

\begin{figure}
\resizebox{0.45\textwidth}{!}{\includegraphics{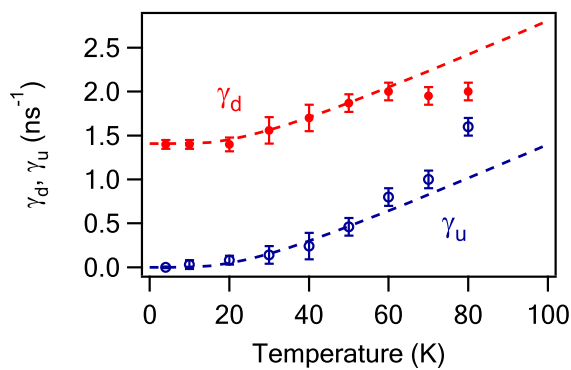}}
\caption{Downward and upward spin flip rates between X and DX as a function of temperature. The dotted lines are the fits using equations (\ref{eq:BES}) and (\ref{eq:gamma}).}
 \label{fig:gammaSP}
\end{figure}

 By fitting the lifetimes using the model of fig \ref{fig:level_schemeD} we can extract
 the values for $\gamma_X $, $\gamma_{NR}$, $\gamma_d$ and $\gamma_u$. Each of these parameters has a specific influence on the shape of the luminescence decay and can be evaluated with a good precision.  The radiative decay rate of the exciton is found to be $\gamma_X=1.4$ ns$^{-1}$. The  values for $\gamma_d$ and $\gamma_u$ are plotted in fig \ref{fig:gammaSP} as a function of temperature. This temperature dependence is well fitted using equations (\ref{eq:BES}) and (\ref{eq:gamma}). This enables us to obtain the DX-X energy splitting $\Delta E=6$ meV and the zero temperature downward rate $\gamma_0=1.4$ ns$^{-1}$.

 The rather large value for the DX-X energy splitting  $\Delta E$ is an indication of the strong exciton confinement within the QD, owing to its relatively small size and the absence of a wetting layer \cite{Puls}. The bulk DX-X energy splitting for CdSe is  $0.12$ meV. According to the calculation performed by Klingshirn et al \cite{Klingshirn}, confinement induced enhancement of this splitting  reaches a factor of $50$ (that is  $\Delta E=6$ meV for CdSe) for infinite barriers cylindrical dots of radius corresponding to the Bohr radius $a_B$ and of height corresponding to $a_B /4$. These dimensions are  compatible with the measured diameter ($10$ nm) of the NW and with the height expected from the CdSe growth duration.
 The  DX-X energy splitting has been measured at $1.9$ meV for self-assembled CdSe/ZnSe QD \cite{Puls}.
 The larger value that we have observed is an indication of the larger confinement in NWs owing to the absence of a wetting layer.
 This $\Delta E=6$ meV splitting corresponds to the value for very small (about 2 nm diameter) colloidal spherical CdSe nanocrystal as reported in reference \cite{Efros}.
The value for zero temperature downward transition rate $\gamma_0=1.4$ ns$^{-1}$ is comparable to what has been obtained for colloidal CdSe nanocrystals \cite{Labeau} or some self assembled InP/GaInP QDs \cite{Reischle}. Slower rates ($\gamma_0=0.01$ ns$^{-1}$)
have also been observed in InGaAs/GaAs self assembled QDs \cite{Smith}.

 A good fitting of the experimental data requires the inclusion of an effective non-radiative  decay rate $\gamma_{NR}$ of the DX state. This rate slightly increases with temperature from $\gamma_{NR}= 0.2$ ns$^{-1}$ at 4K up to $\gamma_{NR}= 0.5$ ns$^{-1}$ at 80 K.
 These values are of similar order of magnitude that what was reported in InGaAs/GaAs self assembled QDs \cite{Smith}.
 Non radiative phenomena in nanocrystals are generally  slower ranging from hundreds of nanoseconds to a few microseconds in colloidal nanocrystals \cite{Labeau,Nirmal}.

 %For these experiments the QD is left in the dark after the light pulse and evolves under no illumination.
 %It can nevertheless be remarked that the  non-radiative  decay rate that we have found for the DX state is comparable %to  the QD charge hopping rate (from neutral to charged QD or vice versa) that we have measured  in reference
 % \cite{correlation,these} and to the spectral diffusion time \cite{these},
 % both obtained under continuous illumination.

To summarize, we have performed temperature dependent lifetime measurements of a CdSe quantum dot embedded in a ZnSe nanowire. A careful quantitative analysis of these data has allowed us to confirm the strong influence of the dark exciton and to extract the dark-bright exciton splitting together with the transition rates between these two  levels.
The rather large dark-bright exciton splitting that we have measured is a signature of the strong confinement of the exciton within the QD in this NW geometry. This value is three times larger than for self-assembled CdSe QDs although the excitonic energy and the X-XX and X-CX  energy splittings are of same order of magnitude.

We thank M. Richard for many stimulating discussions and F. Donatini for very efficient technical support.
T.A. acknowledges support by Deutscher
Akademischer Austauschdienst (DAAD). Part of this
work was supported by European project QAP (Contract No.
15848).

\end{document}